\documentclass[epj]{svjour}
\usepackage{amsmath}
\usepackage{amssymb}
\usepackage[noadjust]{cite}

\newcommand{\aH}[1][]{a^{{#1}\dagger}}
\newcommand{\dip}{\mathrm{d}}
\newcommand{\hatHam}{\hat H}
\newcommand{\hatHem}{\hatHam_{\mathrm{em}}}
\newcommand{\hatHint}{\hatHam_{\mathrm{int}}}
\newcommand{\hatHat}{\hatHam_{\mathrm{at}}}
\newcommand{\locf}[1][\mathrm{diel}]{\mathcal{L}_{#1}}
\newcommand{\relop}[1][]{\mathrm{R}^{#1}}

\newcommand{\kom}[3][]{\left[{#2},{#3}\right]_{\mathrm{#1}}}

\newcommand{\I}{\mathrm{i}}

\newcommand{\bkt}[1]{{\left({#1}\right)}}
\newcommand{\od}[1]{\bkt{#1}}
\newcommand{\abs}[1]{\left|{#1}\right|}

\newcommand{\ket}[1]{\left|{#1}\right>}
\newcommand{\bra}[1]{\left<{#1}\right|}

\begin{document}

\title{Cooperative spontaneous emission in nonuniform media}
\author{Adam Rudzi{\'n}ski\thanks{\email{arudzins@poczta.onet.pl}}}
\institute{Institute of Microelectronics and Optoelectronics, Warsaw University of Technology, Warszawa, Poland}
\date{Received: 5. Aug 2010 / Revised version: 12. Nov 2010}

\abstract{
The subject of this paper is modification of cooperative spontaneous emission by a nonuniform medium,
with nonuniform distributions of electromagnetic field. A brief analyzis is presented and it is postulated,
that if spontaneous emission from an atom is strongly suppressed, cooperative emission with another atom
may be a preferred emission channel and counteract the suppression.
}

\maketitle

\section{Motivation}
It is known, that interaction of an atom with electromagnetic field is affected by the environment
surrounding the atom, whether it is a resonant cavity \cite{Bialynicka96} or, generally, a~nonuniform
medium. In particular, modification of spontaneous emission occurs, what has been first explained by Purcell \cite{Purcell46}.
This fact has found many practical applications, e.g.~following the suggestion of Yablonovitch, that a medium
with a~photonic band gap for all directions of propagation, nowadays called a photonic crystal, could
suppress spontaneous emission \cite{Yablonovitch87}. Another effect concerning spontaneous emission,
known as superradiance, has been predicted by Dicke \cite{Dicke54} and later confirmed in experiments:
systems of many atoms can emit photons cooperatively, what results in radiation intensity proportional
to $N^2$. Obviously, these effects combine, and in result superradiance is modified by the environment as well.
This paper is devoted to analysis of this modification in a nonuniform material, in which
modification of spontaneous emission is significantly different in various regions. Special attention is paid
to the case of an atom is placed in a position in which spontaneous emission is strongly inhibited.
It is postulated, that cooperative emission might be a~preferred decay channel in this case.

\section{Quantum mechanical description of spontaneous emission}
Theory of spontaneous emission is described in many books, e.g. \cite{Meystre91}, and papers, e.g. the one by Glauber and Lewenstein \cite{Glauber91},
where a formulation proper for nonuniform dielectric medium is presented. The most important effects imposed
on spontaneous emission by a material or a structure can be described in a simple non-relativistic approach.
The whole system is decomposed into two subsystems. The first one is the electromagnetic field, described by the hamiltonian
\begin{equation}
\hatHem = \sum_\lambda \hslash \omega_\lambda \hat N_\lambda,
\end{equation}
where $\lambda$ runs through all the modes of the field, $\omega_\lambda$ is angular frequency of the mode $\lambda$
and $\hat N_\lambda$ is the photon number operator for this mode. Eigenstates of $\hat N_\lambda$ form the
Fock basis, which, for spontaneous emission, can be restricted to states with no photons $\ket{0}$ and states
with one photon $\ket{1_\lambda}$. Modes of the structure can form a discrete or
a~continuous (or partly discrete, partly continuous) set, in this latter case summation should be replaced by
integration over proper quantities, represented by $\lambda$ -- this is only a~matter of notation,
for simplicity notation for discrete $\lambda$ will be used here.

The second part of the system is the atomic part, which is assumed to consist of a~number of identical two-level
atoms. For simplicity, any dependence of the environment on atomic states is neglected, assuming identical
excitation energy $\hslash\Omega$ and transition dipole moment $\vec\dip$ for each atom.
State of $N_\mathrm{at}$-atomic system can be defined as tensor product of individual atomic states
\begin{equation}
\ket{\ldots\pm_n\ldots} \equiv \bigotimes_{n=1}^{N_\mathrm{at}} \ket{\pm_n},
\end{equation}
where $\ket{+_n}$ and $\ket{-_n}$ denote the excited and ground state of the $n$-th atom, respectively.
Atomic hamiltonian becomes then
\begin{equation}
\hatHat = \sum_{n=1}^{N_\mathrm{at}} \hslash\Omega \ket{+_n}\bra{+_n}.
\end{equation}

To conveniently describe atomic transitions, one can define a relaxation operator for the $n$-th atom $\relop_n$,
corresponding to deexcitation of this atom:
\begin{eqnarray}
\!\!\!\!\relop_n\ket{\ldots +_n\ldots} = \vec\dip \ket{\ldots -_n\ldots}, &\ & \relop_n\ket{\ldots -_n\ldots} = 0.
\end{eqnarray}
Then, its hermitean conjugate $\relop[\dagger]_n$ corresponds to putting the $n$-th atom in the excited state.
The interaction hamiltonian allowing to describe spontaneous emission can be written in the form:
\begin{equation}
\hatHint = -\hslash \sum_\lambda \sum_{n=1}^{N_\mathrm{at}} g_{\lambda,n} \bkt{ \I \relop[\dagger]_n \cdot \vec f_\lambda\od{\vec r_n} a_\lambda + \mathrm{h.c.} },
\end{equation}
where $\vec r_n$ is the position of the $n$-th atom, $\vec f_\lambda\od{\vec r}$ is the electric field distribution
of mode $\lambda$, abbreviation $\mathrm{h.c.}$ stands for hermitean conjugate of the previous term, and
\begin{equation}
g_{\lambda,n} \equiv g_\lambda\od{\vec r_n} = \locf\od{\vec r_n} \frac{\Omega}{\sqrt{2\hslash\varepsilon_0\omega_\lambda}} .
\end{equation}
The factor $\locf\od{\vec r}$ describes relation between amplitudes of local and macroscopic electric field \cite{Glauber91, deVries98},
which may be position dependent. $a_\lambda$ is the anihilation operator for mode $\lambda$, which satisfies bosonic
commutation rules with its hermitean conjugate, the creation operator $\aH_\lambda$:
\begin{eqnarray}
&& \kom{a_\lambda}{\aH_\mu}=\delta_{\lambda\mu} ,\\
&& \kom{a_\lambda}{a_\mu}=\kom{\aH_\lambda}{\aH_\mu}=0.
\end{eqnarray}

Let the system be initially in a state $\ket{0}\ket{\Psi_0}$, where $\ket{\Psi_0}$ is a state with $N$
excited atoms. Because coupling of the atomic system with electromagnetic field is weak, it is reasonable
to consider only the states, which couple directly to $\ket{0}\ket{\Psi_0}$, i.e. these, which combine
into $\hatHint\ket{0}\ket{\Psi_0}$, which are states with one photon and $N-1$ excited atoms.
All the other states can be produced by coupling of higher order, several magnitudes smaller, and will be neglected.
This approximation allows to define the Hilbert space for the considered problem and conduct derivation
of spontaneous emission rate the same way, as for a relatively simple single atomic case \cite{Meystre91}.
State vector of the system can be defined as:
\begin{multline}
\ket{\psi\od{t}} =
C_0\od{t} \mathrm{e}^{-\I N\Omega t} \ket{0}\ket{\Psi_0}
\\
+
\sum_\lambda \sum_{\Psi'} C_{\Psi'\lambda}\od{t} \mathrm{e}^{-\I \left[\bkt{N-1}\Omega + \omega_\lambda \right] t} \ket{1_\lambda}\ket{\Psi'},
\end{multline}
where $\Psi'$ runs through states with $N-1$ excited atoms. Initial conditions relevant for spontaneous emission
are: $C_0\od{0}=1$ and the remaining amplitudes $C_{\Psi'\lambda}\od{0}=0$.
Projecting the Schr{\"o}dinger equation onto individual states from the state vector one obtains a set of equations of motion for amplitudes $C_0$
and $C_{\Psi'\lambda}$, which are most elegantly treated by application of Laplace transformation,
leading to a set of linear equations for their Laplace transforms. This way, one easily obtains a formal
solution for $C_0$:
\begin{equation}
C_0\od{t} = \mathcal{L}^{-1}\left\{ \tilde C_0\od{s} \right\},
\end{equation}
with
\begin{multline}
\tilde C_0\od{s} =
\\
\bkt{s + \frac{1}{\hslash^2}\sum_\lambda\sum_{\Psi'} \frac{\abs{\bra{1_\lambda}\bra{\Psi'} \hatHint \ket{0}\ket{\Psi_0}}^2}{ s-\I\bkt{\Omega-\omega_\lambda} }}^{-1},
\end{multline}
where $\mathcal{L}^{-1}$ stands for the inverse Laplace transform. The solution can be easily written
as a function of time in Weisskopf-Wigner approximation, which is obtained by evaluating the second term
in the nominator of $\tilde C\od{s}$ in the limit $s\rightarrow 0^+$. This procedure leads to the well-known result
\begin{equation}
\abs{C_0\od{t}}^2 = \mathrm{e}^{-\Gamma t},
\end{equation}
with the total decay rate
\begin{equation}
\Gamma = \sum_\lambda \Gamma_\lambda \delta\od{\Omega-\omega_\lambda}
\end{equation}
and rate of emission into mode $\lambda$ given by:
\begin{multline}
\Gamma_\lambda =
\\
2\pi \sum_{n=1}^{N_\mathrm{at}} g_{\lambda,n} \sum_{n'=1}^{N_\mathrm{at}} g_{\lambda,n'} \left< \relop[\dagger]_n \cdot \vec f_\lambda\od{\vec r_n} \relop_{n'} \cdot \vec f^*_\lambda\od{\vec r_{n'}} \right>,\label{eq:Gamma}
\end{multline}
with $\left< \cdot \right>$ denoting mean value in the state $\ket{\Psi_0}$.
This is a~standard derivation, but its reconstruction helps to precisely define the considered problem.

\section{Cooperative spontaneous emission}
The result \eqref{eq:Gamma} for the emission rate explains existence of the most important phenomena.
Except for description of the simplest case of only one atom, it allows to observe the term proportional
to $N^2$ and it reveals dependence of the decay on distribution of electric field. Thus, it is sufficient
for an analyzis of spontaneous emission decay in structures, where distributions $\vec f_\lambda$ depend strongly on position, what is the case of interest in this paper.

Let us split the expression \eqref{eq:Gamma} into two terms:
\begin{equation}
\Gamma_\lambda = \sum_n \Gamma^{(1)}_\lambda\od{n} + \sum_n \sum_{n'\neq n} \Gamma^{(2)}_\lambda\od{n,n'},
\end{equation}
the first term describing contributions of individual atoms:
\begin{equation}
\Gamma^{(1)}_\lambda\od{n} = 2\pi g_{\lambda,n}^2 \left< \relop[\dagger]_n \cdot \vec f_\lambda\od{\vec r_n} \relop_n \cdot \vec f^*_\lambda\od{\vec r_n} \right>
\end{equation}
and the second one describing contributions of atomic pairs:
\begin{multline}
\Gamma^{(2)}_\lambda\od{n,n'} =
\\
2\pi g_{\lambda,n} g_{\lambda,n'} \left< \relop[\dagger]_n \cdot \vec f_\lambda\od{\vec r_n} \relop_{n'} \cdot \vec f^*_\lambda\od{\vec r_{n'}} \right>.
\end{multline}
These expressions allow to study how field distributions affect the decay rate. Contributions of individual atoms
are modified accordingly to $\abs{\vec f_\lambda\od{\vec r_n}}^2$. In particular, if no modes with angular frequency
$\Omega$ are present in the structure, emission at this frequency is completely stopped. This is the idea
which has been presented by Yablonovitch \cite{Yablonovitch87}. However, this is impossible in real structures,
which are of finite size and there always exists a set of modes with a~given frequency. It is possible though, that
a~structure very strongly extinguishes field of a mode in a certain region. Therefore, strong inhibition
of spontaneous emission into a mode can occur, but because the structure adequately forms the field distribution,
and not because this distribution is not present at all.

This fact may have considerable consequences for collective emission by atomic pairs, which is described by
the term $\Gamma^{(2)}_\lambda\od{n,n'}$. If both atoms of a pair are placed nearby, they interact with
modal fields having approximately the same intensities at their positions, thus, modification of emission
is similar to single-atomic case and, apart from superradiant emission \cite{Dicke54}, no new effect is observed.
However, if the atoms are placed in positions, at which intensity of a mode $\abs{\vec f_\lambda}$ differs significantly
(this might occur e.g. in a finite one-dimensional photonic crystal for a mode from a photonic band gap, for which
field distribution is strongly suppressed in the middle of the structure, but outside forms a partly standing wave
with much larger amplitude \cite{APPA111}), rate of cooperative emission by the pair into the mode is proportional
to product of these intensities, significantly different than their second powers.
Thus, for atoms $n$-th and $n'$-th, if $\abs{\vec f_\lambda\od{\vec r_n}} \ll \abs{\vec f_\lambda\od{\vec r_{n'}}}$,
emission rate from the $n$-th atom into the mode $\Gamma^{(1)}_\lambda\od{n}$ is proportional to $\abs{\vec f_\lambda\od{\vec r_n}}^2$
and can be significantly suppressed, but the factor related to electromagnetic field in the rate of collective emission
by the pair $\Gamma^{(2)}_\lambda\od{n,n'}$ is much bigger $\abs{\vec f_\lambda\od{\vec r_n}}\abs{\vec f_\lambda\od{\vec r_{n'}}}$.
This means, that electromagnetic field strongly privileges this emission channel and makes emission
more probable.

For two atoms to emit a photon cooperatively a correlation (at least partial entanglement) between their states is necessary.
This correlation may appear beacause of radiative processes \cite{Guo06, Yang08}, thus cooperative
emission may occur even if atoms are pumped incoherently. Another restriction on cooperative emission is
the existence of a~maximal distance, at which the atom can emit in cooperation: radiation
emited by one of the atoms must reach the second atom before the act of emission is finished.
This condition for atoms in vacuum has been discussed by Arecchi and Courtens \cite{Arecchi70}.
They have argued, that if the lifetime of the excited state is $\tau_\mathrm{ex}$, then the critical
distance is simply $L_c = c\tau_\mathrm{ex}$ and have provided a formula, which can be written in the following form:
\begin{equation}
L_c = \frac{2}{\lambda_0}\sqrt{\frac{c}{\mathcal{N} \Gamma_1}},\label{eq:Lc}
\end{equation}
where $\lambda_0$ is the wavelength of the emitted radiation, $\mathcal{N}$ -- concentration of radiating atoms
and $\Gamma_1$ -- emission rate for a single atom. For optical transitions in gases $L_c$ is
of the order of $10\ \mathrm{cm}$. Formula \eqref{eq:Lc} can be used also for a rough estimation of $L_c$
in a~structure. Let us assume $\lambda_0\approx 1\ \mathrm{\mu m}$, concentration $\mathcal{N}\approx 10^{19}\ \mathrm{cm}^{-3}$,
and a~value typical for optical transitions $\Gamma_1\approx 10^9\ \mathrm{s}^{-1}$. Finally, to conform to
the idea of the condition, instead of $c$ one should use $v_\mathrm{g}$, group velocity of light in the structure.
This leads to the result $L_c \approx 3 \sqrt{v_\mathrm{g}/c} \times 10^{-1}\ \mathrm{m}$. Even if the group velocity
in the structure dropped to a few $\mathrm{m/s}$, i.e. $v_\mathrm{g}/c\approx 10^{-8}$, it would lead to
$L_c\approx 30\ \mathrm{\mu m}$. More realistic values can be found in literature, where observations of slow light
with $v_\mathrm{g}/c\approx 10^{-2}$ are reported \cite{Altug05, Vlasov05, Kawasaki07} and this ratio
can be used as a~representative value.
Then, one obtains $L_c \approx 3\ \mathrm{cm}$, a~distance longer, than dimensions of optoelectronics
devices. Therefore, the described phenomenon of collective emission into a mode, for which emission from
a~single atom would be strongly suppresed, could become a dominant process and might be observed
in structures, where strong inhibition of spontaneous emission occurs. This fact would have practical
consequences, because it would counteract stopping the emission or enhancing the lifetime
of atomic excited state by a structure strongly modifying distributions of electromagnetic field.

Physical systems are more complicated than just isolated ensembles of atoms and, in general, interactions
of different kind than direct radiative coupling between the atoms are present. These interactions can
also cause collective states to appear and contribute to superradiant emission this way. For example,
recently existence of superradiance in a system with plasmonic coupling has been discussed \cite{Pustovit09}.
A similar effect of structure nonuniformity should be observable in these systems as well.

This work has been financially supported by State Committee for Scientific Research, project N N515 052535.

\end{document}